\def\nk{n_{\rm b}}

\def\Pb{P_{\rm b}}

\def\rfr#1{Equation\,(\ref{#1})}
\def\rfrs#1#2{Equations\,(\ref{#1})\,to\,(\ref{#2})}

\def\virg#1{``#1"}

\def\eqi{\begin{equation}}
\def\eqf{\end{equation}}
\def\eqia{\begin{eqnarray}}
\def\eqfa{\end{eqnarray}}

\def\rp#1#2{{#1\over#2}}
\def\lb#1{\label{#1}}

\def\bds#1{\boldsymbol{#1}}

\def\ton#1{\left(#1\right)}
\def\qua#1{\left[#1\right]}
\def\grf#1{\left\{#1\right\}}

\documentclass[onecolumn]{aastex}
\usepackage{morefloats}
\usepackage[title]{appendix}
\usepackage{hyperref}
\usepackage{booktabs}
\usepackage[table,xcdraw]{xcolor}
\usepackage{multirow}
\usepackage{rotating,tabularx}
\usepackage{float}
\usepackage{enumerate}
\usepackage{rotating}
\usepackage[polutonikogreek,english]{babel}
\usepackage{amsmath,starfont,textgreek,w-greek,wasysym}
\usepackage{amsthm}
\usepackage{amscd,lineno}
\usepackage{amssymb,dsfont}
\usepackage{graphicx,epsfig}
\usepackage{txfonts}
\usepackage{xr-hyper}
\usepackage{breakurl}
\usepackage{url}

\RequirePackage{color}

\newcommand{\emaila}{lorenzo.iorio@libero.it}

\allowdisplaybreaks[1]

\begin{document}

\title{A comment on \virg{Lense–Thirring frame dragging induced by a fast-rotating white dwarf in a binary pulsar system} by V. Venkatraman Krishnan \textit{et al.}}

\shortauthors{L. Iorio}

\author{Lorenzo Iorio\altaffilmark{1} }
\affil{Ministero dell'Istruzione, dell'Universit\`{a} e della Ricerca
(M.I.U.R.)-Istruzione
\\ Permanent address for correspondence: Viale Unit\`{a} di Italia 68, 70125, Bari (BA),
Italy}

\email{\emaila}

\begin{abstract}
We comment on a recent study reporting evidence for the general relativistic Lense-Thirring secular precession of the inclination $I$ of the orbital plane to the plane of the sky of the tight binary system PSR J1141-6545 made of a white dwarf and an emitting radiopulsar of comparable masses. The quadrupole mass moment $Q_2^\mathrm{c}$ and  the angular momentum ${\bds S}^\mathrm{c}$ of the white dwarf cause the detectable effects on $I$ with respect to the present-day accuracy in the pulsar's timing. The history-dependent and model-dependent assumptions to be made on $Q_2^\mathrm{c}$ and ${\bds S}^\mathrm{c}$,
required even just to calculate the analytical expressions for the resulting post-Keplerian precessions, may be deemed as too wide in order to claim a successful test of the Einsteinian gravitomagnetic effect. Moreover, depending on how $Q_2^\mathrm{c}$ is calculated, the competing quadrupole-induced rate of change, which is a major source of systematic uncertainty, may be up to $\lesssim 30-50\,\mathrm{per\,cent}$ of the Lense-Thirring effect for most of the allowed values in the 3D parameter space spanned by the white dwarf's spin period $P_\mathrm{s}$, and the polar angles $i_\mathrm{c},\,\zeta_\mathrm{c}$ of its spin axis. The possible use of the longitude of periastron $\varpi$ is investigated as well. It turns out that a measurement of its secular precession, caused, among other things, also by $Q_2^\mathrm{c},\,{\bds{S}}^\mathrm{c}$, could help in further restricting the permitted regions in the white dwarf's parameter space.
\end{abstract}


keywords{
gravitation-stars: pulsars: general-stars: white dwarfs
}

\section{Introduction}
Recently, \citet{LTWDPSR20} claimed a successful detection of the general relativistic Lense-Thirring (LT) precession of the inclination of the orbital plane to the plane of the sky in the tight full two-body system PSR J1141-6545 \citep{2000ApJ...543..321K,2011MNRAS.412..580A} hosting an emitting radiopulsar p with mass $M_\mathrm{p}=1.27\,\mathrm{M}_\odot$ whose companion c is a massive white dwarf (WD) with $M_\mathrm{c}=1.02\,\mathrm{M}_\odot$.
The aforementioned Einsteinian effect belongs to a wide class of phenomena which general relativity\footnote{See, e.g., \citet{2016Univ....2...23D} and references therein for a comprehensive overview of that theory and of the challenges it currently faces} predicts they arise from mass-energy currents \citep{1986SvPhU..29..215D,2002NCimB.117..743R,2004GReGr..36.2223S,2009SSRv..148...37S}. In the case of, say, a localized astronomical rotating source, the latter ones constitute the body's spin dipole moment, i.e. its proper angular momentum $\bds S$. Because of the formal resemblance, occurring in the slow-motion and weak-field limit, of the linearized approximation of the Einstein field equations with the linear Maxwellian equations of electromagnetism, such a phenomenology is collectively dubbed as \virg{gravitomagnetism} \citep{Thorne86,1986hmac.book..103T,1988nznf.conf..573T}, despite it has nothing to do with the magnetic fields and the electric currents. Steady experimental efforts lead to the successful measurement of another gravitomagnetic effect some years ago, i.e. the precession of the spin of an orbiting gyroscope in the field of a twisting body \citep{Pugh59,Schiff60}, with the spaceborne Gravity Probe B (GP-B) mission around the Earth \citep{2015CQGra..32v4001E}. The final accuracy was $19\,\mathrm{per\,cent}$, contrary to the $\simeq 1\,\mathrm{per\,cent}$ level initially expected \citep{2001LNP...562...52E}. For other ongoing or proposed attempts with natural or non-dedicated artificial satellites of major astronomical bodies in the Solar System, see, e.g., \citet{2011Ap&SS.331..351I,2013CEJPh..11..531R,2015grmb.book..125C,Lucchesi019}, and references therein.
It may be, at this point, the case to note that putting the GP-B dedicated experiment and the attempts with the Earth's artificial satellites of the LAGEOS family and the Satellite Laser Ranging (SLR) technique on the same foot does not correspond to the actual state of affairs. Indeed, while the GP-B results have not yet been criticized so far in any published paper in the peer-reviewed literature, the SLR-based attempts by Ciufolini and coworkers have been so far the subject of a staggering number of published peer-reviewed papers by some authors criticizing them; see, e.g., \citet{2013CEJPh..11..531R}, and references therein.
The measurement of the LT periastron precession  in the double pulsar PSR J0737-3039 \citep{2003Natur.426..531B,2004Sci...303.1153L}, composed by two neutron stars, is actively pursued as well \citep{2018mgm..conf.1860K}.

\citet{LTWDPSR20} made use of the measurement
\eqi
\dot x_\mathrm{p}^\mathrm{exp}\pm\upsigma_{\dot x_\mathrm{p}^\mathrm{exp}}=\ton{1.7\pm 0.3}\times 10^{-13}\,\mathrm{s\,s}^{-1}\lb{mesu}
\eqf
of the secular change $\dot x_\mathrm{p}$ of the pulsar's projected semimajor axis
\eqi
x_\mathrm{p} = \rp{a_\mathrm{p}}{c}\,\sin I.\lb{xp}
\eqf
In \rfr{xp}, $c$ is the speed of light in vacuum, $a_\mathrm{p}$ is the barycentric semimajor axis of the pulsar p, and $I$ is the inclination of the binary's orbital plane to the plane of the sky or, equivalently, of the system's orbital angular momentum $\bds L$ to the line of sight.

\citet{LTWDPSR20} correctly argued that the dynamical part $\dot x_\mathrm{p}^\mathrm{dyn}$ of $\dot x_\mathrm{p}$, able to explain about $79\,\mathrm{per\,cent}$ of \rfr{mesu}, comes from the rate of change of the inclination $I$, so that
\eqi
\dot x_\mathrm{p}^\mathrm{dyn} = \rp{a_\mathrm{p}}{c}\,\cos I\,\dot I = x_\mathrm{p}\,\cot I\,\dot I\lb{xdyn}.
\eqf
Indeed, there are certain post-Keplerian (PK) dynamical features of a full two-body system made of comparable masses $M_\mathrm{A},\,M_\mathrm{B}$ like just PSR J1141-6545 which, under certain circumstances, can induce a secular change of $I$. They are the quadrupole mass moments $Q_2^\mathrm{A},\,Q_2^\mathrm{B}$, causing a Newtonian PK acceleration, and the spin angular momenta ${\bds S}_\mathrm{A},\,{\bds S}_\mathrm{B}$, responsible of a general relativistic PK acceleration which, in the limiting case of a test particle orbiting a fixed primary, reduces to the LT one \citep{1918PhyZ...19..156L,Sof89,1991ercm.book.....B,SoffelHan19}. The explicit expressions for the secular precessions of the angle $\mathfrak{I}$ between the orbital plane and an arbitrary reference $\grf{x,\,y}$ plane induced by such physical effects can be found in, e.g., \citet{2017EPJC...77..439I}. To the benefit of the reader, we display them here\footnote{As far as \rfr{ILT} is concerned, see also, e.g., \citet[Equation\,(3.27)]{1992PhRvD..45.1840D}.}:
\begin{align}
\dot{\mathfrak{I}}_{Q_2} \lb{IQ2}& = \rp{3\,\nk}{2\,p^2}\qua{\rp{Q^\mathrm{A}_2}{M_\mathrm{A}}\ton{{\bds{\hat{S}}}^\mathrm{A}\bds\cdot\bds{\hat{l}}}
\ton{{\bds{\hat{S}}}^\mathrm{A}\bds\cdot\bds{\hat{\nu}}} + \rp{Q^\mathrm{B}_2}{M_\mathrm{B}}\ton{{\bds{\hat{S}}}^\mathrm{B}\bds\cdot\bds{\hat{l}}}
\ton{{\bds{\hat{S}}}^\mathrm{B}\bds\cdot\bds{\hat{\nu}}}}, \\ \nonumber \\
\dot{\mathfrak{I}}_\mathrm{LT} \lb{ILT} & = \rp{2\,G}{c^2\,a^3\,\ton{1-e^2}^{3/2}}\grf{\qua{\ton{1+\rp{3}{4}\rp{M_\mathrm{B}}{M_\mathrm{A}}}{\bds S}^\mathrm{A} + \ton{1+\rp{3}{4}\rp{M_\mathrm{A}}{M_\mathrm{B}}}{\bds S}^\mathrm{B}}\bds\cdot\bds{\hat{l}}},
\end{align}
where $G$ is the Newtonian constant of gravitation, $a$ is the semimajor axis of the relative orbit, $e$ is the orbital eccentricity, $p\doteq a\,\ton{1-e^2}$ is the semilatus rectum, $\nk\doteq\sqrt{G\ton{M_\mathrm{A}+M_\mathrm{B}}/a^3}$ is the Keplerian mean motion, $\bds{\hat{l}}\doteq\grf{\cos\Omega,\,\sin\Omega,\,0}$ is the unit vector of the line of the nodes pointing towards the ascending node, $\Omega$ is the longitude of the ascending node locating the position of the orbital plane in the adopted reference $\grf{x,\,y}$ plane, $\bds{\hat{\nu}}\doteq\grf{\sin \mathfrak{I}\,\sin\Omega,\,-\sin\mathfrak{I}\,\cos\Omega,\,\cos\mathfrak{I}}$ is the out-of-plane unit vector directed along the orbital angular momentum $\bds L$ tilted by the angle $\mathfrak{I}$ to the chosen reference $z$ axis.
In the following analysis, we will direct the $z$ axis along the line of sight in such a way that the plane of the sky will be our $\grf{x,\,y}$ reference plane, and $\mathfrak{I}=I$. The spin axes of the two bodies $\mathrm{A},\,\mathrm{B}$ will be parameterized as follows
\begin{align}
{\hat{S}}^\mathrm{A/B}_x & = \sin i_\mathrm{A/B}\,\cos\varphi_\mathrm{A,B},\\ \nonumber \\
{\hat{S}}^\mathrm{A/B}_y & = \sin i_\mathrm{A/B}\,\sin\varphi_\mathrm{A,B},\\ \nonumber \\
{\hat{S}}^\mathrm{A/B}_z & = \cos i_\mathrm{A/B},
\end{align}
so that if $i_\mathrm{A/B}=90\,\mathrm{deg}$, the spin axis of A or B lies in the plane of the sky.
However, it turns out that the azimuthal angles $\varphi_\mathrm{A/B}$ enter \rfrs{IQ2}{ILT} always in the form $\zeta_\mathrm{A/B}\doteq \varphi_\mathrm{A/B}-\Omega$, which are the angles of the spin axes' projections onto the plane of the sky reckoned from the (unknown) line of the nodes.
It should be noted that, in general astronomical and astrophysical scenarios, $i_\mathrm{A/B}\,\zeta_\mathrm{A/B}$ are unknown for one or both the binary's components $\mathrm{A},\,\mathrm{B}$. Instead, in the Earth-satellite scenario, $i_\oplus=0$ since the orientation of the terrestrial angular momentum in space is known, so that the reference $z$ axis is usually chosen to be aligned just with ${\bds S}_\oplus$ at some reference epoch.
In the following, A will denote the pulsar p, while B is the WD c.

It should be noted that, in addition to the masses, the eccentricity and the orbital period\footnote{In turn, knowing $\Pb$ allows to extract the relative semimajor axis via the Third Kepler Law.} $\Pb$ which are all determined from the timing analysis, \rfrs{IQ2}{ILT} contain nine additional parameters which, in principle, must be known for a test of the PN gravitomagnetic LT effect which, among other things, would be unavoidably biased, to an extent which has to be assessed as accurately as possible, by the Newtonian quadrupolar field: the magnitudes of both the quadrupole mass moments $Q_2^\mathrm{c},\,Q_2^\mathrm{p}$ and of the angular momenta $S^\mathrm{c},\,S^\mathrm{p}$, the two pairs of angles fixing the directions of the two bodies' spin axes $\bds{\hat{S}}^\mathrm{c},\,\bds{\hat{S}}^\mathrm{p}$ in space, and the longitude of the ascending node $\Omega$. A test of the LT effect necessarily implies the knowledge of all such key physical and orbital parameters along with their associated uncertainties, to be obtained in an independent way from, say, the measurement of other PK parameters and/or their rates, which is not the case here. Even if one wanted to perform a preliminary sensitivity analysis aimed to evaluate the perspectives of measuring frame-dragging with some uncertainty, to be assessed as well, in face of the current level of accuracy in experimentally determining the orbital effects which are supposedly impacted by the gravitomagnetic field, some a-priori guesses about, say, the quadrupoles and the angular momenta from some physical models of the structure of the bodies involved along with their uncertainties are needed because they enter in the analytical formulas for the precessions of interest.
Conversely, one can a priori assume the validity of general relativity (and of the Newtonian quadrupole-induced dynamics as well), and use it to try to constrain (some of) the system's parameters; in fact, it is the route practically followed by \citet{LTWDPSR20}. In this case, it may be misleading presenting their own results as a test of the LT effect, or as a demonstration of its existence, also because a quantitative assessment of the total error budget, including known major sources of systematic uncertainty like the WD's quadrupole mass moment, should be released.
%
%

It turns out that, in fact, the number of the relevant, a-priori unknown parameters is less than nine. Indeed, in Section\,\ref{psr}, it will be shown that both the PK rates of change induced by the pulsar's $Q_2^\mathrm{p}$ and ${\bds S}^\mathrm{p}$ can be neglected because, for virtually all plausible values of their key parameters, they are smaller $\upsigma_{\dot x_\mathrm{p}^\mathrm{exp}}$. Moreover, as already remarked, the longitude of the ascending node $\Omega$, which is not measurable with usual timing analysis because, usually, it is not present in the timing formula, enters the analytical expressions of \rfrs{IQ2}{ILT} in the form $\zeta_\mathrm{c}\doteq \varphi_\mathrm{c}-\Omega_\mathrm{c}$. That reduces the needed parameters to $Q_2^\mathrm{c},\,S_\mathrm{c},\,i_\mathrm{c},\,\zeta_\mathrm{c}$. As we will see in Sections\,\ref{QI}-\ref{ploti}, one can make some a-priori assumptions on the WD's quadrupole and moment of inertia-affected by unavoidable and non-negligible uncertainties-and look at the WD's spin period and orientation angles as parameters to be constrained by imposing certain conditions.
%
%
%
\section{The quadrupole mass moment and the angular momentum of the pulsar}\lb{psr}
According to \citet{1999ApJ...512..282L}, for a neutron star we have
\eqi
Q_2^\mathrm{p} = \xi_\mathrm{p}\,\rp{M_\mathrm{p}^3\,G^2}{c^4},\lb{Qpsr}
\eqf
where $\left|\xi_\mathrm{p}\right|$ ranges from $0.074$ to $3.507$ for a variety of Equations of State (EOSs) and $M_\mathrm{p}=1.4\,\mathrm{M}_\odot$; cfr. Table\,4 of \citet{1999ApJ...512..282L}.
The maximum value $\left|\xi_\mathrm{p}\right|\simeq 3.507$, inserted in \rfr{Qpsr}, yields for the pulsar PSR J1141–6545
\eqi
Q_2^\mathrm{p}\lesssim 3.1\times 10^{37}\,\mathrm{kg\,m^2}.\lb{QPS}
\eqf
As we will see in Section\,\ref{QMM}, \rfr{QPS} is much smaller than the quadrupole mass moment of the WD.

For the angular momentum of the pulsar, \citet[pag.\,580]{LTWDPSR20} proposed a maximum value
\eqi
S^\mathrm{p} = 4\times 10^{40}\,\mathrm{kg\,m^2\,s^{-1}}\lb{SPS}
\eqf
for the angular momentum of a recycled pulsar.

By inserting \rfrs{QPS}{SPS} in \rfrs{IQ2}{ILT}, it is possible to find the maximum (absolute) value of the part of $\dot x_\mathrm{p}^{\mathrm{PK}}$ due to the pulsar only with respect to $i_\mathrm{p},\,\zeta_\mathrm{p}$: one gets
\eqi
\left|\dot x_\mathrm{p}^\mathrm{PK}\right|^\mathrm{p}\leq 2.8\times 10^{-14}\,\mathrm{s\,s^{-1}},
\eqf
which, in fact, is slightly smaller than $\upsigma_{\dot x_\mathrm{p}^\mathrm{exp}}$.
Thus, in the following we will consider only the angular momentum and the quadrupole mass moment of the WD.
\section{The uncertainties in the WD's angular momentum and quadrupole mass moment}\lb{QI}
If the measurement of $\dot x_\mathrm{p}$ has to be interpreted as a genuine test of the LT effect, both $S^\mathrm{c}$ and  ${\bds{\hat{S}}}^\mathrm{c}$ should be known independently of $\dot x_\mathrm{p}$  itself. Moreover, also the accuracy with which $Q_2^\mathrm{c}$ is known should be stated in order to assess the impact of its uncertainty on the alleged relativistic test since the former should be regarded as a major source of systematic uncertainty for the latter. Conversely, even if one assumes the validity of the PK effects under consideration to constrain, say, the WD's spin period $P_\mathrm{c}$ and the angles $i_\mathrm{c},\,\zeta_\mathrm{c}$ fixing the orientation in space of its spin axis, some guesses about the WD's moment of inertia $\mathcal{I}_\mathrm{c}$ and $Q_2^\mathrm{c}$ are needed in order to have a manageable parameter space.

Several model-dependent assumptions driven by the composition and evolutionary history of the WD must be made both on $\mathcal{I}_\mathrm{c}$ and $Q_2^\mathrm{c}$. As we will see in the next Sections, \citet{LTWDPSR20} essentially relied upon \citet{2017MNRAS.464.4349B} for such key physical parameters of the WD. \citet{2017MNRAS.464.4349B} studied the equilibrium configurations of uniformly rotating WDs using the Chandrasekhar \citep{1931ApJ....74...81C,1939isss.book.....C} and Salpeter \citep{1961ApJ...134..683H,1961ApJ...134..669S} EoSs at zero temperature in the framework of Newtonian physics. However, one should not ignore that, since the pioneering work by Chandrasekhar, considerable progress has been achieved in the determination of the EoS of dense Coulomb plasmas. Such EoSs, which are far more realistic than that by Chandrasekhar, can be very easily implemented numerically since analytic fits exist, as described, e.g., in \citet{NSbook2007}, and various codes are publicly available. In particular, fits for hot dense Coulomb liquid plasmas, as found in hot WDs, have been recently presented in \citet{2019MNRAS.490.5839B}; see also references therein. General relativistic treatments of rotating WDs can be found, e.g., in \citet{1971Ap......7..274A,Boshkayev_2012}.

For the sake of definiteness, in the following we will follow  \citet{2017MNRAS.464.4349B} in order to infer our own evaluations of the relevant physical parameters of the WD.
\subsection{The uncertainties in the WD's moment of inertia}
On the one hand,  \citet{LTWDPSR20} did not provide any estimate for $\mathcal{I}_\mathrm{c}$ at pag. 11 of their Supplementary Materials citing \citet{1992PhRvD..45.1840D} which, actually, did not deal with such a physical parameter of WDs at all. On the other hand, they suggested at pag. 14 of their Supplementary Materials
\eqi
\mathcal{I}_\mathrm{c} \simeq 0.9\times 10^{43}\,\mathrm{kg\,m}^2\lb{boh?}
\eqf
invoking Equation\,(4) of \citet{2017MNRAS.464.4349B}. Apart from the fact that it is unclear from  \citet{LTWDPSR20} if \rfr{boh?} refers to the present-day state of the WD or to its initial  configuration at the beginning of the accretion phase, Equation\,(4) of \citet{2017MNRAS.464.4349B} is valid only for a non-rotating, static body.
Instead, from\footnote{In it, the units of the normalizing factor $I^\ast$ of the moment of inertia are mistakenly $\mathrm{g\,cm}^3$.} Figure\,5 of \citet{2017MNRAS.464.4349B}, plotting the moment of inertia as a function of the mass for static and rotating WDs, it seems that, for $M_\mathrm{WD}\simeq 1\,\mathrm{M}_\odot$, $\mathcal{I}_\mathrm{WD}\simeq 1-3\times 10^{43}\,\mathrm{kg\,m}^2$ depending on the WD's composition. The dotted red curve of Figure\,5 of \citet{2017MNRAS.464.4349B} for a rotating $^4$He WD like the one of interest here seems to point towards\footnote{Such a value is close to the value obtainable for a solid sphere $\mathcal{I}_\mathrm{c}=2/5 M_\mathrm{c} R_\mathrm{c}^2=2.36\times 10^{43}\,\mathrm{kg\,m}^2$ calculated for $R_\mathrm{c}=5400\,\mathrm{km}$ \citep{LTWDPSR20}.} $\mathcal{I}_\mathrm{WD}\simeq 2\times 10^{43}\,\mathrm{kg\,m}^2$.
In particular, a subtle issue in computing $S^\mathrm{c}$ consists of the fact that using the product of the moment of inertia times the spin frequency  holds only if one adopts the moment of inertia $\mathcal{I}^{(0)}$ for the \textit{static}, i.e. \textit{nonrotating} configuration of the WD, as per Equation\,(16) of \citet{2017MNRAS.464.4349B}. From the continuous red curve of Figure\,5 of \citet{2017MNRAS.464.4349B} for a static $^4$He WD, one may infer $\mathcal{I}_\mathrm{c}^{(0)}\simeq 1\times 10^{43}\,\mathrm{kg\,m}^2$ for $M_\mathrm{c}=1.02\,\mathrm{M}_\odot$. On the other hand, such a value for the WD's mass refers to what actually measured by \citep{LTWDPSR20}, not to the static mass which would be required for $\mathcal{I}^{(0)}$.
\subsection{Uncertainties in the WD's quadrupole moment}\lb{QMM}
Concerning the quadrupole  $Q_2$ of a body of mass $M$ and radius $R$ having dimensional units of $\textrm{kg\,m}^2$, it can be expressed in terms of $R$ and of the corresponding dimensionless  moment $J_2$ ($J_2>0$ for an oblate body) as
\eqi
Q_2 = -J_2\,M\,R^2,\lb{zumzum}
\eqf
where the Newtonian formula \citep{2009ApJ...698.1778R}
\eqi
J_2 = \rp{k_2}{3}\,q,\lb{Quo}
\eqf
is often adopted for a wide range of weakly relativistic astrophysical objects like gaseous giant planets, main-sequence stars, and  WDs as well \citep{2017RAA....17...61M,2017MNRAS.464.4349B}. In \rfr{Quo}, it is
\eqi
q\doteq \rp{4\uppi^2}{P^2}\,\rp{R^3}{GM},
\eqf
where $P$ is the body's spinning period,
and $k_2$ is its Love number \citep{1939MNRAS..99..451S}. The latter can be thought of as a measure of the level of central condensation of an object, with stronger central condensation corresponding to smaller $k_2$ \citep{2009ApJ...698.1778R};
for main-sequence stars, it is $k_2\simeq 0.03$ \citep{1995A&AS..109..441C}.

\citet{LTWDPSR20} claimed to have inferred the value of $k_2$ by means of the results in \citet{2017MNRAS.464.4349B}. Let us check the finding of \citet{LTWDPSR20} by attempting to recover our own value for the Love number relying upon \citet{2017MNRAS.464.4349B}.

By looking at the dashed red curves for a rotating $^4$He WD in Figure\,1, Figure\,9 and Figure\,14 of \citet{2017MNRAS.464.4349B}, which are all computed with $q=1$, it can be inferred\footnote{In Figure\,14 of \citet{2017MNRAS.464.4349B}, the units of $Q^\ast$ are mistakenly reported as $\mathrm{g\,cm}^3$. } $Q_2^\mathrm{c}\simeq 4.5\times 10^{42}\,\mathrm{kg\,m}^2$ for our WD. By using such a figure in \rfrs{zumzum}{Quo}, calculated with $M=1.02\,\mathrm{M}_\odot,\,R=5400\,\mathrm{km}$ \citep{LTWDPSR20} and $q=1$, one gets
\eqi
k^\mathrm{c}_2=0.228.\lb{Love}
\eqf

It should be remarked that \rfr{Love} may not be straightforwardly compared with the value $k_2=0.081$ reported in \citet{LTWDPSR20}. Indeed, \rfr{xdyn} and \rfr{IQ2} yield
\eqi
\dot x^{Q_2^\mathrm{c}}_\mathrm{p} = -\rp{x_\mathrm{p}\,\cot I\,\nk\,k^\mathrm{c}_2\,q_\mathrm{c}\,R_\mathrm{c}^2}{2a^2\,\ton{1-e^2}^2}\,F\ton{i_\mathrm{c},\,\zeta_\mathrm{c},\,I},\lb{iQ}
\eqf
where $F$ is a certain function, not displayed explicitly here, of the inclination $I$ of the binary's orbital plane and of the angles $i_\mathrm{c},\,\zeta_\mathrm{c}$ characterizing the orientation in space of the WD' spin axis.
\rfr{iQ} agrees with Equation\,(S7) of \citet{LTWDPSR20}, apart from the sign and the definition of the function $F$ since \citet{LTWDPSR20} adopted a different parameterization for ${\bds{\hat{S}}}^\mathrm{c}$; this fact suggests that, implicitly, also \citet{LTWDPSR20} adopted \rfrs{zumzum}{Quo} to model the WD's quadrupole moment. Now,
\citet{LTWDPSR20}, defined a dimensionless, (positive) quadrupole mass moment in their equation\,(S8) given, in our notation\footnote{For \citet{LTWDPSR20}, it is $I\rightarrow i,\,q_\mathrm{c}\rightarrow{\hat{\Omega}}^2_\mathrm{c}$.}, by
\eqi
Q\doteq\rp{k^\mathrm{c}_2\,R_\mathrm{c}^2\,q_\mathrm{c}}{2\,a^2\,\ton{1-e^2}^2}.\lb{ballo}
\eqf
\citet{LTWDPSR20} claimed to have used the equation-of-state and composition
independent $I$-Love-$Q$ relations by \citet{2017MNRAS.464.4349B} in order to infer
\eqi
k_2^\mathrm{c}=0.081.\lb{k2V}
\eqf
Now,
it is unclear how \citet{LTWDPSR20} may, actually, have used the results by \citet{2017MNRAS.464.4349B} to obtain \rfr{k2V}. Indeed, even by assuming that also \citet{LTWDPSR20} computed \rfr{ballo} with $q_\mathrm{c}=1$ to subsequently compare it with some of the curves in \citet{2017MNRAS.464.4349B}, the dimensionless quadrpole moment $\overline{Q}$ of \citet{2017MNRAS.464.4349B} has nothing to do with \rfr{ballo}, being, instead, defined as $\overline{Q}\doteq c^2\,Q\,M/S^2$. Be that as it may, \citet{2017MNRAS.464.4349B} did not deal at all with binary systems hosting a WD; as such, neither the semimajor axis $a$ nor the orbital eccentricity $e$ enter any of the formulas for $Q$ by \citet{2017MNRAS.464.4349B}, contrary to the definition of \rfr{ballo}.

About the calculation of $Q_2^\mathrm{c}$, there is also the following subtle issue pertaining the WD's quantities $M_\mathrm{c},\,R_\mathrm{c}$ entering \rfr{zumzum} which may make uncertain the previous evaluation(s) of the Love number $k_2$.
\citet{LTWDPSR20}, without any apparent justification, proposed $R_\mathrm{c}\simeq 5400\,\mathrm{km}$ for their value $M_\mathrm{c}=1.02\,\mathrm{M}_\odot$ of the WD's mass; seemingly, \citet{LTWDPSR20} assumed only that the WD is made of an ideal degenerate Fermi gas by citing \citet{10.1093/mnras/91.5.456} which, however, did not provide any estimates of WD's radii.
The quantity $R$ entering the formulas for $Q_2$ displayed so far is \textit{not} the equatorial radius of the body under consideration assumed to be \textit{rotating}; instead, it has to be meant as the radius of its \textit{static} configuration, as explicitly pointed out in \citet{2017MNRAS.464.4349B}. Now, since the equatorial radius for a static case reduces to the static radius, using the continuous red curve for a static $^4$He WD in Figure\,2 of \citet{2017MNRAS.464.4349B},
in fact, one would infer $R_\mathrm{c}\simeq 5400\, \mathrm{km}$ for $M_\mathrm{c}=1.02\,\mathrm{M}_\odot$. But such a value for the WD's mass, which corresponds to what actually measured by \citet{LTWDPSR20},  can only refer to the \textit{total}, i.e. \textit{rotating} mass of the WD at hand. The dashed red curve for a \textit{rotating} $^4$He WD in Figure\,2 of \citet{2017MNRAS.464.4349B} yields an equatorial radius as large as $R_\mathrm{c}^\mathrm{eq}=10000\,\mathrm{km}$ for $M_\mathrm{c}/\mathrm{M}_\odot=1.02$. The same problem arises also for the correct value of the WD's mass to be used in the formulas for $Q_2^\mathrm{c}$ since it must refer to the \textit{static} configuration as well \citep{2017MNRAS.464.4349B}, while \citet{LTWDPSR20} had experimentally access only to the \textit{total}, i.e. the \textit{rotating} mass of the WD. Unfortunately, it does not seem that there is a way to sort out the static values of the WD's mass and radius from \citet{2017MNRAS.464.4349B} which would be required to correctly compute the WD's quadrupole mass moment $Q_2^\mathrm{c}$ by means of \rfrs{zumzum}{Quo}.
%
%
%
%
\section{Constraints on the WD's spin period and spin axis'orientation from $\dot x_\mathrm{obs}$}\lb{ploti}
Here, we will look at the sum $\dot x_\mathrm{p}^\mathrm{PK}\doteq \dot x_\mathrm{p}^\mathrm{LT}+\dot x_\mathrm{p}^{Q_2}$ of the dynamical PK secular rates of change of the pulsar's projected semimajor axis due to the PN gravitomagnetic field and the Newtonian quadrupole of the WD as a function of 3 independent variables: the WD's spin period $P_\mathrm{c}$, and the two angles $i_\mathrm{c},\,\zeta_\mathrm{c}$ determining the orientation of the WD's spin axis in space. Then, we will impose the condition that $\dot x^\mathrm{PK}_\mathrm{p}$ lies within a certain interval  of the experimental range for $\dot x_\mathrm{p}$, and will inspect the resulting constraints on $P_\mathrm{c},\,i_\mathrm{c},\,\zeta_\mathrm{c}$. We will not make any a priori assumptions on $i_\mathrm{c},\,\zeta_\mathrm{c}$, and  allow $P_\mathrm{c}$ to vary just from the minimum value to avoid centrifugal breakup $P^\mathrm{min}_\mathrm{c}=7\,\mathrm{s}$ \citep{LTWDPSR20} to $P^\mathrm{max}_\mathrm{c}=1\,\mathrm{hr}=3600\,\mathrm{s}$, which is close to the spin period ($P=1.13\,\mathrm{hr}$) of the fastest rotating isolated WD known (SDSS J0837+1856) having mass $M\simeq 0.9\,\mathrm{M}_\odot$ similar to $M_\mathrm{c}$ \citep{2017ApJ...841L...2H}.

\begin{figure}[htb]
\begin{center}
\centerline{
\vbox{
\begin{tabular}{c}
\epsfysize= 10.0 cm\epsfbox{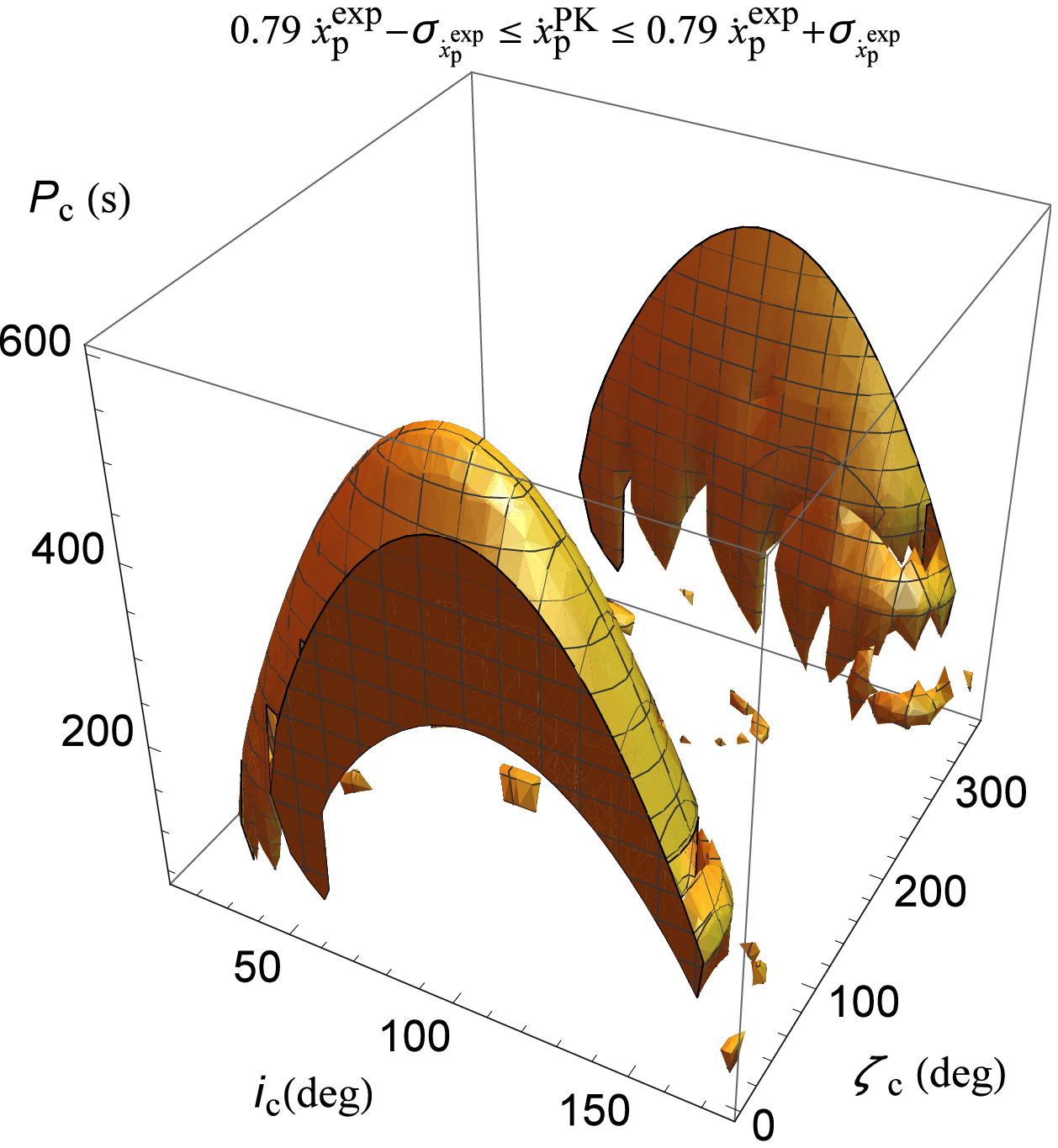}\\
\end{tabular}
}
}
\caption{Allowed regions in the 3D parameter space $\grf{i_\mathrm{c},\,\zeta_\mathrm{c},\,P_\mathrm{c}}$ determined by the constraint $ 0.79\,\dot x_\mathrm{p}^\mathrm{exp}-\upsigma_{\dot x_\mathrm{p}^\mathrm{exp}}\leq \dot x_\mathrm{p}^\mathrm{PK}\ton{i_\mathrm{c},\,\zeta_\mathrm{c},\,P_\mathrm{c}}\leq 0.79\,\dot x_\mathrm{p}^\mathrm{exp}+\upsigma_{\dot x_\mathrm{p}^\mathrm{exp}}$. Each point on the coloured surface corresponds to a set of values of $i_\mathrm{c},\,\zeta_\mathrm{c},\,P_\mathrm{c}$ which allow $\dot x_\mathrm{p}^\mathrm{PK}\ton{i_\mathrm{c},\,\zeta_\mathrm{c},\,P_\mathrm{c}}$ to lie within $0.79\,\dot x_\mathrm{p}^\mathrm{exp}\pm \upsigma_{\dot x_\mathrm{p}^\mathrm{exp}}$. In calculating $\dot x_\mathrm{p}^\mathrm{PK}$, both the Newtonian quadrupolar and the LT precessions of $I$ were simultaneously taken into account by using $k_2^\mathrm{c}=0.228,\,M_\mathrm{c}=1.02\,\mathrm{M}_\odot,\,R_\mathrm{c}=5400\,\mathrm{km},\,\mathcal{I}_\mathrm{c}^{\ton{0}}=1\times 10^{43}\,\mathrm{kg\, m^2}$.}\label{figura1}
\end{center}
\end{figure}

%
Figure\,\ref{figura1} shows the allowed regions, in colour, of the 3D parameter space $\grf{i_\mathrm{c},\,\zeta_\mathrm{c},\,P_\mathrm{c}}$ determined by the condition
\eqi
0.79\,\dot x_\mathrm{p}^\mathrm{exp}-\upsigma_{\dot x_\mathrm{p}^\mathrm{exp}}\leq \dot x_\mathrm{p}^\mathrm{PK}\ton{i_\mathrm{c},\,\zeta_\mathrm{c},\,P_\mathrm{c}}\leq \dot 0.79\,x_\mathrm{p}^\mathrm{exp}+\upsigma_{\dot x_\mathrm{p}^\mathrm{exp}}.\lb{boundo}
\eqf
We adopted the values $k_2^\mathrm{c}=0.228,\,M_\mathrm{c}=1.02\,\mathrm{M}_\odot,\,R_\mathrm{c}=5400\,\mathrm{km},\,\mathcal{I}_\mathrm{c}^{\ton{0}}=1\times 10^{43}\,\mathrm{kg\, m^2}$. It can be noted that the WD's spin period $P_\mathrm{c}$ cannot be larger than $\simeq 600\,\mathrm{s}$; on the other hand, such a value can occur only for very few values of $\zeta_\mathrm{c}$ and, especially, $i_\mathrm{c}$. In general, $\zeta_\mathrm{c}$ appears to be more effectively constrained than $i_\mathrm{c}$ since there are three relatively narrow, disjointed intervals of allowed values for it.

\begin{figure}[htb]
\begin{center}
\centerline{
\vbox{
\begin{tabular}{cc}
\epsfysize= 7.0 cm\epsfbox{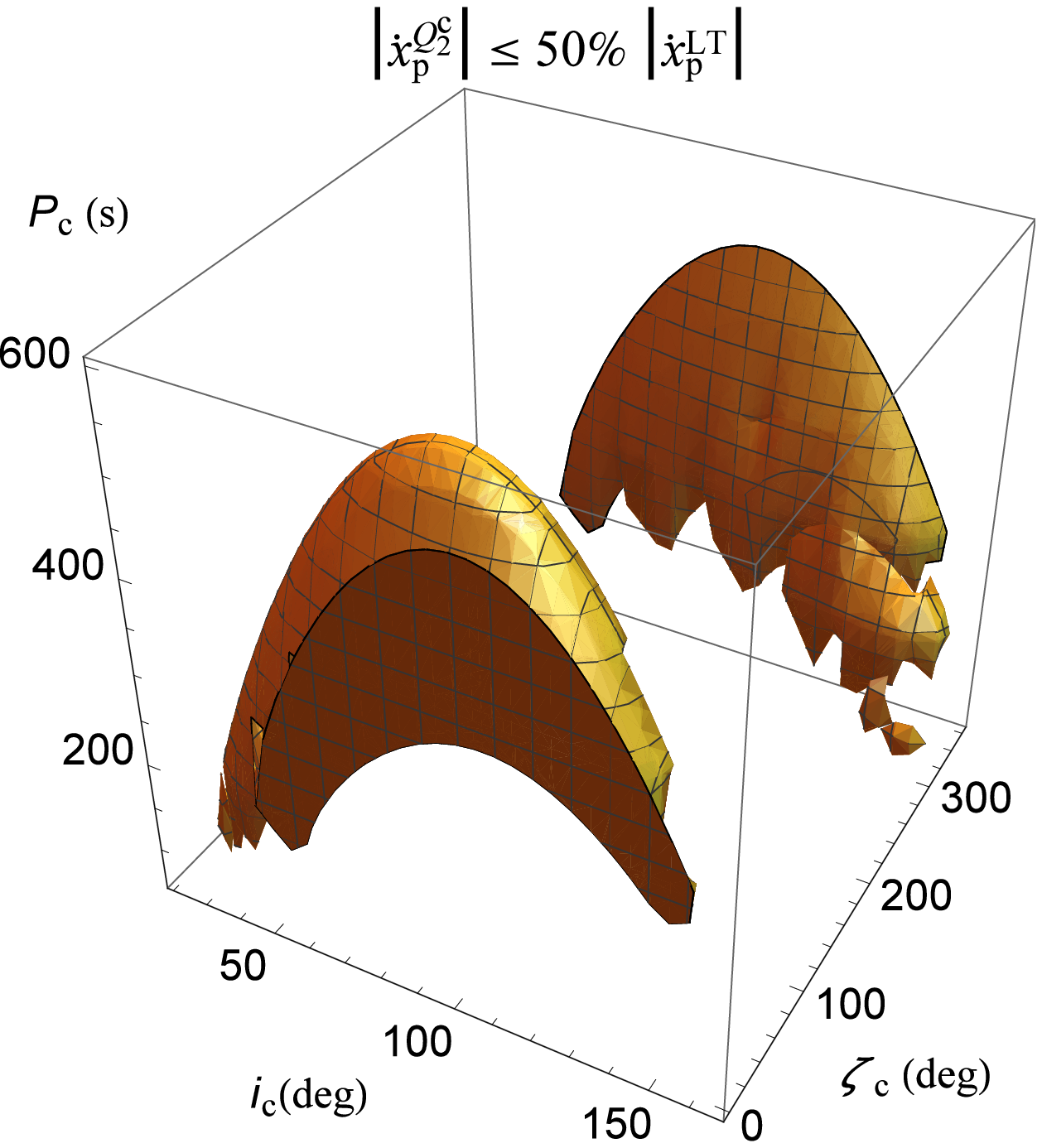}& \epsfysize= 7.0 cm\epsfbox{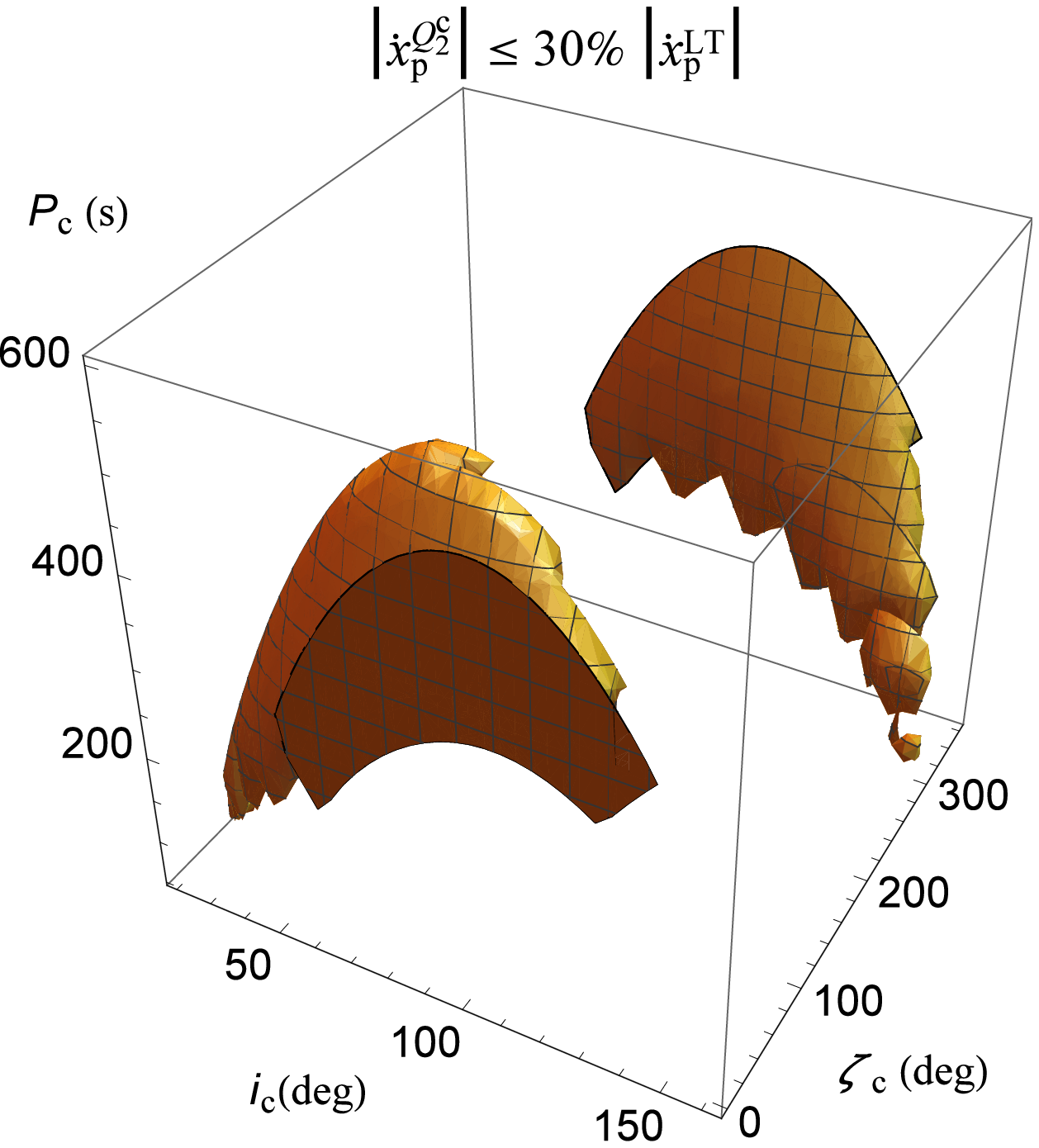}\\
\epsfysize= 7.0 cm\epsfbox{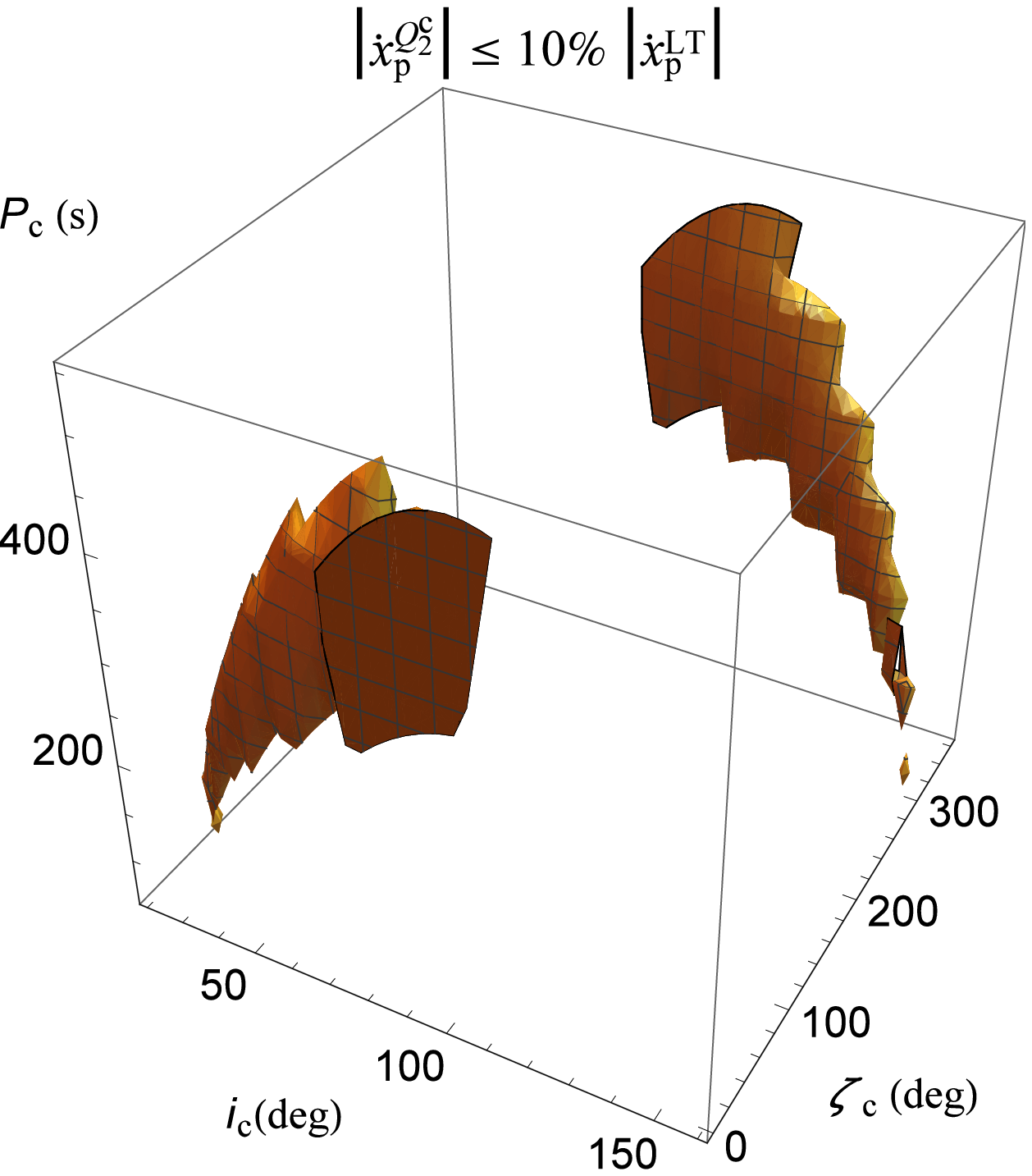}& \epsfysize= 7.0 cm\epsfbox{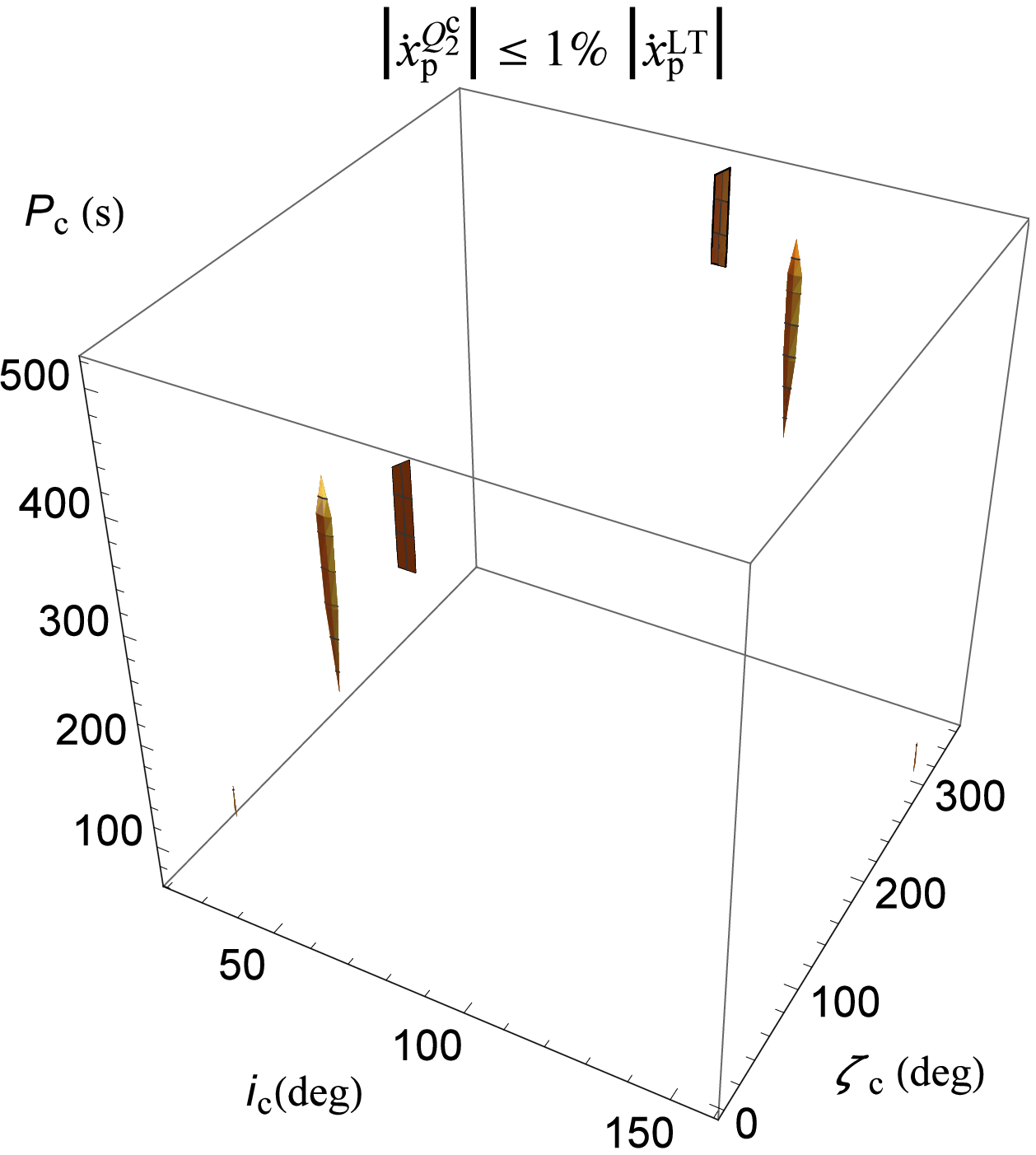}\\
\end{tabular}
}
}
\caption{Allowed regions in the 3D parameter space $\grf{i_\mathrm{c},\,\zeta_\mathrm{c},\,P_\mathrm{c}}$ determined simultaneously by the constraints $ 0.79\,\dot x_\mathrm{p}^\mathrm{exp}-\upsigma_{\dot x_\mathrm{p}^\mathrm{exp}}\leq \dot x_\mathrm{p}^\mathrm{PK}\ton{i_\mathrm{c},\,\zeta_\mathrm{c},\,P_\mathrm{c}}\leq 0.79\,\dot x_\mathrm{p}^\mathrm{exp}+\upsigma_{\dot x_\mathrm{p}^\mathrm{exp}}$ and $\left|\dot x^{Q_2^\mathrm{c}}_\mathrm{p}/\dot x^\mathrm{LT}_\mathrm{p}\right|\leq \mathrm{X}\,\,\mathrm{per\,cent}$ with $\mathrm{X}=50$ (right upper panel), $30$ (left upper panel), $10$ (right lower panel), $1$ (left lower panel). Each point on the coloured surfaces corresponds to a set of values of $i_\mathrm{c},\,\zeta_\mathrm{c},\,P_\mathrm{c}$ which allow $\dot x_\mathrm{p}^\mathrm{PK}\ton{i_\mathrm{c},\,\zeta_\mathrm{c},\,P_\mathrm{c}}$ to lie within $0.79\,\dot x_\mathrm{p}^\mathrm{exp}\pm \upsigma_{\dot x_\mathrm{p}^\mathrm{exp}}$ in such a way that the Newtonian quadrupolar effect amounts to $\mathrm{X}\,\,\mathrm{per\,cent}$ of the PN LT one. The values $k_2^\mathrm{c}=0.228,\,M_\mathrm{c}=1.02\,\mathrm{M}_\odot,\,R_\mathrm{c}=5400\,\mathrm{km},\,\mathcal{I}_\mathrm{c}^{\ton{0}}=1\times 10^{43}\,\mathrm{kg\, m^2}$ were used.}\label{figura2}
\end{center}
\end{figure}

In order to evaluate the impact of the competing quadrupole rate on the general relativistic effect, viewed as a source of systematic error on the latter,  we plot the allowed regions determined simultaneously by the condition of \rfr{boundo} and by the bound
\eqi
\left|\rp{\dot x_\mathrm{p}^{Q_2^\mathrm{c}}}{\dot x_\mathrm{p}^\mathrm{LT}}\right|\leq \mathrm{X}\,\mathrm{per\,cent},
\eqf
where X ranges from 1 to 50, in Figure\,\ref{figura2}.
It can be noted that it is rather unlikely that the systematic error due to the WD's oblateness on the LT rate can be as little as $\lesssim 1-10\,\mathrm{per\,cent}$ since such a condition would occur only for a very limited set of values in the system's 3D parameter space. On the contrary, it is much more likely that the quadrupolar rate can be as large as $\lesssim 30-50\,\mathrm{per\,cent}$ of its relativistic counterpart. Also Figure\,\ref{figura2} was obtained by using the same numerical values for the system's key parameters as Figure\,\ref{figura1}.
\section{Using the periastron precession?}
\citet{LTWDPSR20} estimated the periastron precession $\dot\omega_\mathrm{exp}$ with an uncertainty $\upsigma_{\dot\omega_\mathrm{exp}}$, not publicly released for some reasons, larger than the sum of both the PK precessions of interest. Thus, \citet{LTWDPSR20}  concluded that no useful constraints on the system's parameters could be inferred from the measured periastron precession.
On the other hand, in a private exchange with the present author, V. Venkatraman Krishnan told him that, in fact, he and his groups did not determine the periastron precession separately. Instead, they computed the well known PN gravitoelectric formula
\eqi
\dot\omega^\mathrm{PN}_\mathrm{GE} =\rp{3\,\nk\,\mu}{c^2\,a\,\ton{1-e^2}}\lb{schwa}
\eqf
for the PN periastron precession due to two mass monopoles,
expressed it in terms of the experimentally measured masses, eccentricity and orbital period, and propagated their uncertainties presumably obtaining
\eqi
\upsigma_{\dot\omega_\mathrm{GE}^\mathrm{PN}}= 9.3\times 10^{-5}\,\mathrm{deg\,yr^{-1}}=5.1\times 10^{-14}\,\mathrm{s^{-1}}.\lb{errpge}
\eqf
In fact, \rfr{errpge} may not be viewed as the actual experimental error $\upsigma_{\dot\omega_\mathrm{exp}}$ of a purely phenomenological, model-independent determination of $\dot\omega_\mathrm{exp}$ as an additional PK parameter. Instead, \rfr{errpge} can only be deemed as the present-day uncertainty on the PN gravitoelectric periastron precession induced by two pointlike masses due to the current errors in the system's parameters $\mu,\,e,\,\Pb$.

The topic of the periastron precession is potentially an important one  since, as it will be shown below, the claim by \citet{LTWDPSR20} about the  uselessness of $\dot\omega_\mathrm{exp}$ may be relaxed, depending on the size of $\upsigma_{\dot\omega_\mathrm{exp}}$ and on the accurate modeling of the periastron precessions of interest.

Before proceeding further, we, first, note that a potentially relevant ambiguity may arise since \citet{LTWDPSR20} used the symbol $\omega$ for a Keplerian orbital element generically identified as \virg{periastron} throughout their paper. Actually, in the literature, $\omega$ is customarily adopted to designate the argument of pericentre, which is an angle lying in the orbital plane reckoned from the line of the nodes to the point of closest approach along the orbit. On the other hand, in their Table\,1, \citet{LTWDPSR20} used the same symbol to indicate the longitude of periastron which, instead, is a broken angle usually defined as $\varpi\doteq\Omega+\omega$. As long as one considers only the PN gravitoelectric field due to two mass monopoles, then there is no difference between $\omega$ and $\varpi$ because $\dot\Omega^\mathrm{PN}_\mathrm{GE} =0$.
When, however, one considers the PK quadrupolar and LT effects,  it is, then, important to distinguish between $\omega$ and $\varpi$ because, in general, both $\omega$ and $\Omega$ undergo secular precessions due to $Q_2$ and $\bds S$.
To the benefit of the reader, we report here from \citet{2017EPJC...77..439I} the general expressions for the quadrupolar and LT secular precessions of the argument of pericentre $\omega$
\begin{align}
-\rp{4\,p^2}{3\,\nk}\,\dot\omega_{Q_2} \nonumber & = \rp{Q_2^\mathrm{A}}{M_\mathrm{A}}\grf{2 - 3\qua{\ton{\bds{\hat{S}}^\mathrm{A}\bds\cdot\bds{\hat{l}}}^2 + \ton{\bds{\hat{S}}^\mathrm{A}\bds\cdot\bds{\hat{m}}}^2} + 2\ton{\bds{\hat{S}}^\mathrm{A}\bds\cdot\bds{\hat{m}}}\,\ton{\bds{\hat{S}}^\mathrm{A}\bds\cdot\bds{\hat{\nu}}}\,\cot I}+\\ \nonumber \\
&+\rp{Q_2^\mathrm{B}}{M_\mathrm{B}}\grf{2 - 3\qua{\ton{\bds{\hat{S}}^\mathrm{B}\bds\cdot\bds{\hat{l}}}^2 + \ton{\bds{\hat{S}}^\mathrm{B}\bds\cdot\bds{\hat{m}}}^2} + 2\ton{\bds{\hat{S}}^\mathrm{B}\bds\cdot\bds{\hat{m}}}\,\ton{\bds{\hat{S}}^\mathrm{B}\bds\cdot\bds{\hat{\nu}}}\,\cot I}, \\ \nonumber \\
\dot\omega_\mathrm{LT} & = -\rp{2\,G}{c^2\,a^3\,\ton{1-e^2}^{3/2}}\qua{\ton{1+\rp{3}{4}\rp{M_\mathrm{B}}{M_\mathrm{A}}}{\bds S}^\mathrm{A} + \ton{1+\rp{3}{4}\rp{M_\mathrm{A}}{M_\mathrm{B}}}{\bds S}^\mathrm{B}}\bds\cdot\ton{2\,\bds{\hat{\nu}}+\cot I\,\bds{\hat{m}}},
\end{align}
where $\bds{\hat{m}}=\grf{-\cos\mathfrak{I}\,\sin\Omega,\,\cos\mathfrak{I}\,\cos\Omega,\,\sin\mathfrak{I}}$ is a unit vector
directed transversely to the line of the nodes in the orbital plane,
and of the longitude of the ascending node $\Omega$
\begin{align}
\dot\Omega_{Q_2} \lb{OQ2}& = \rp{3\,\nk\,\csc I}{2\,p^2}\qua{\rp{Q^\mathrm{A}_2}{M_\mathrm{A}}\ton{{\bds{\hat{S}}}^\mathrm{A}\bds\cdot\bds{\hat{m}}}
\ton{{\bds{\hat{S}}}^\mathrm{A}\bds\cdot\bds{\hat{\nu}}} + \rp{Q^\mathrm{B}_2}{M_\mathrm{B}}\ton{{\bds{\hat{S}}}^\mathrm{B}\bds\cdot\bds{\hat{m}}}
\ton{{\bds{\hat{S}}}^\mathrm{B}\bds\cdot\bds{\hat{\nu}}}}, \\ \nonumber \\
\dot\Omega_\mathrm{LT} \lb{OLT}& =\rp{2\,G\,\csc I}{c^2\,a^3\,\ton{1-e^2}^{3/2}}\,\qua{\ton{1+\rp{3}{4}\rp{M_\mathrm{B}}{M_\mathrm{A}}}{\bds S}^\mathrm{A} + \ton{1+\rp{3}{4}\rp{M_\mathrm{A}}{M_\mathrm{B}}}{\bds S}^\mathrm{B}}\bds\cdot\bds{\hat{m}}.
\end{align}
In view of the fact that the actual position of the line of the nodes is unmeasurable for PSR J11416545,  we are inclined to use the longitude of periastron $\varpi$.
Now, \citet{1998MNRAS.298...67W}, cited by \citet{LTWDPSR20} at pag. 12 of their Supplementary materials, claimed to look, among other things, at the longitude of the pericentre\footnote{For \citet{1998MNRAS.298...67W} it is $\mathfrak{I}\rightarrow\theta,\Omega\rightarrow\phi,\,\varpi\,(\mathrm{or}\,\omega\,?)\rightarrow\psi$.} $\varpi$, but the Lagrange planetary equation allegedly for it displayed in Equation\,(34) of \citet{1998MNRAS.298...67W} is, actually, the one for the argument of pericentre $\omega$ \citep{2011rcms.book.....K,2014grav.book.....P}. Moreover, apart from the fact that \citet{1998MNRAS.298...67W} worked in the test particle limit-which, however, in the present case may be viewed as a minor issue because only the spin and the quadrupole of the WD do matter-a major issue is certainly represented by the fact that \citet{1998MNRAS.298...67W} did not consider the PN LT effect at all since he dealt only with the quadrupole of the primary of the binary\footnote{It is a binary pulsar whose companion is a main-sequence star.} considered in that particular study. Last but not least, Equation\,(40) in \citet{1998MNRAS.298...67W} is the well known averaged shift per orbit of the (argument of) pericentre in the standard Earth-satellite case when the spin axis is aligned with the reference $z$ axis. Thus, the results by \citet{1998MNRAS.298...67W}, if really used by \citet{LTWDPSR20}, may not represent a firm ground for an analysis of the role of the PK quadrupolar and periastron precessions in the binary system at hand.

The next step consists of trying to figure out what could plausibly be the experimental uncertainty in the experimentally measured precession of the (longitude of) periastron from the data publicly released by \citet{LTWDPSR20}. A possible way  may be taking the ratio of the known error
\eqi
\upsigma_{\varpi_\mathrm{exp}}=6\times 10^{-4}\,\mathrm{deg}
\eqf
in the determination of the (longitude of) periastron at epoch displayed in Table\,1 of  \citet{LTWDPSR20} to the overall time span of the data analysis $\Delta t=18.03\,\mathrm{yr}$. In doing that, we are assuming that, in their Table\,1, \citet{LTWDPSR20} actually referred to the longitude of periastron $\varpi$ instead of the argument of periastron $\omega$. A scaling factor $\kappa$, calibrated with the the rate of the projected semimajor axis,  may be applied as well since, as we will show below, taking straightforwardly the ratio of the experimental uncertainty in an orbital parameter to the data span may  yield somewhat optimistic figures.
Since it is
\eqi
\rp{\upsigma_{x_\mathrm{p}}}{\Delta t}=\rp{3\times 10^{-6}\,\mathrm{s}}{18.03\,\mathrm{yr}=5.7\times 10^8\,\mathrm{s}}=5.3\times 10^{-15}\,\mathrm{s\,s^{-1}},
\eqf
a comparison with \citet{LTWDPSR20}
\eqi
\upsigma_{\dot x_\mathrm{p}}=3\times\times 10^{-14}\,\mathrm{s\,s^{-1}}
\eqf
yields a scaling factor of
\eqi
\kappa = \rp{\Delta t\,\upsigma_{\dot x_\mathrm{p}}}{\upsigma_{x_\mathrm{p}}}=5.7.
\eqf
Thus, we tentatively infer
\eqi
\upsigma_{\dot\varpi_\mathrm{exp}}\simeq \kappa\rp{\upsigma_{\varpi_\mathrm{exp}}}{\Delta t}=1.9\times 10^{-4}\,\mathrm{deg\,yr^{-1}}=1.04\times 10^{-13}\,\mathrm{s^{-1}}.\lb{erro}
\eqf
\begin{figure}[htb]
\begin{center}
\centerline{
\vbox{
\begin{tabular}{c}
\epsfysize= 10.0 cm\epsfbox{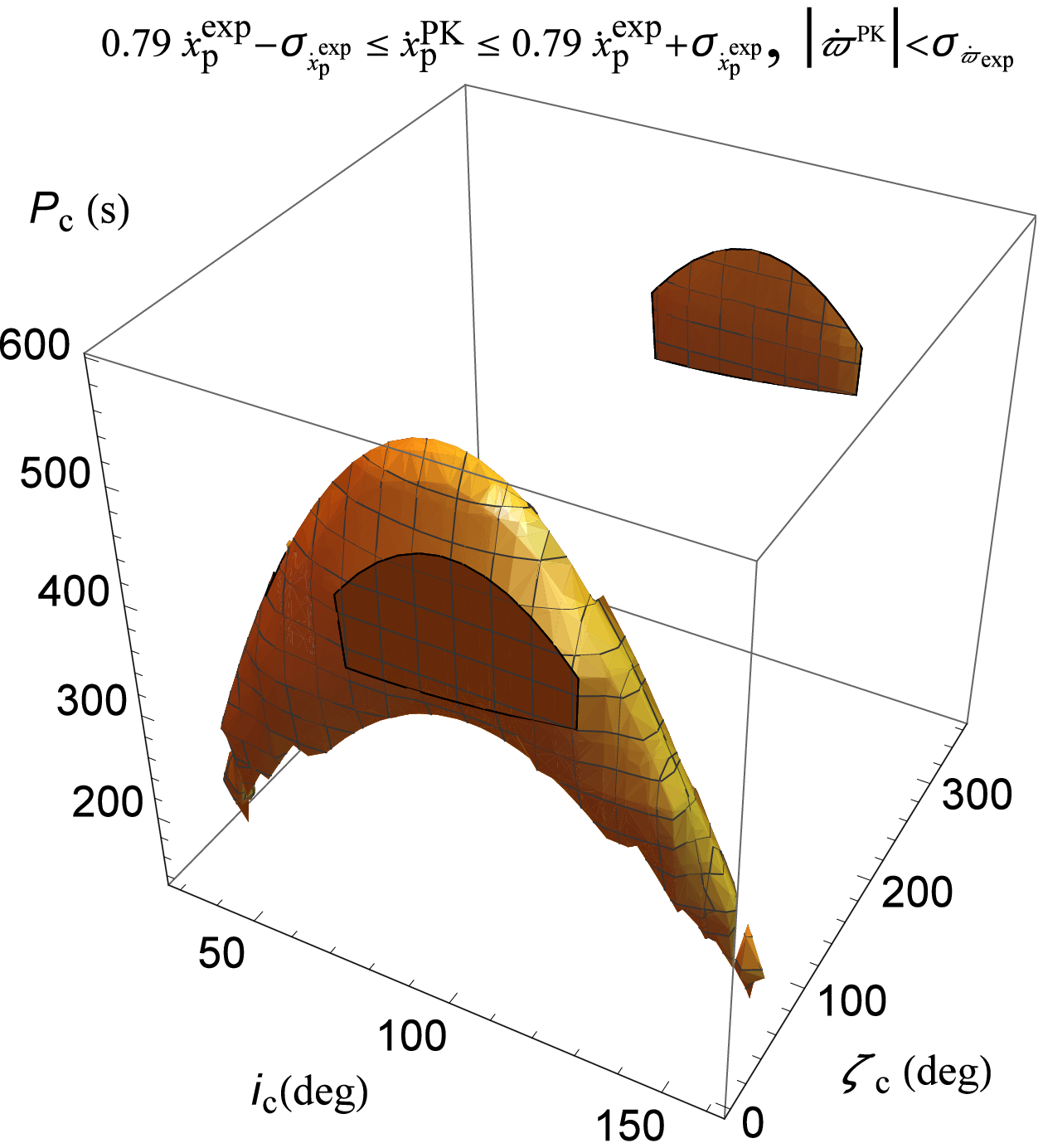}\\
\end{tabular}
}
}
\caption{Allowed region in the 3D parameter space $\grf{i_\mathrm{c},\,\zeta_\mathrm{c},\,P_\mathrm{c}}$ determined simultaneously by the constraints $ 0.79\,\dot x_\mathrm{p}^\mathrm{exp}-\upsigma_{\dot x_\mathrm{p}^\mathrm{exp}}\leq \dot x_\mathrm{p}^\mathrm{PK}\ton{i_\mathrm{c},\,\zeta_\mathrm{c},\,P_\mathrm{c}}\leq 0.79\,\dot x_\mathrm{p}^\mathrm{exp}+\upsigma_{\dot x_\mathrm{p}^\mathrm{exp}}$, and $\left|\dot\varpi^\mathrm{PK}\right|<\upsigma_{\dot\varpi_\mathrm{exp}}$, with $\upsigma_{\dot\varpi_\mathrm{exp}} \simeq 1.9\times 10^{-4}\,\mathrm{deg\,yr^{-1}}$, as per \rfr{erro}. Each point on the coloured surface corresponds to a set of values of $i_\mathrm{c},\,\zeta_\mathrm{c},\,P_\mathrm{c}$ which allow simultaneously $\dot x_\mathrm{p}^\mathrm{PK}\ton{i_\mathrm{c},\,\zeta_\mathrm{c},\,P_\mathrm{c}}$ to lie within $0.79\,\dot x_\mathrm{p}^\mathrm{exp}\pm \upsigma_{\dot x_\mathrm{p}^\mathrm{exp}}$ and $\left|\dot\varpi^\mathrm{PK}\right|<\upsigma_{\dot\varpi_\mathrm{exp}}$. In calculating $\dot x_\mathrm{p}^\mathrm{PK},\,\dot\varpi^\mathrm{PK}$, both the Newtonian quadrupolar and the LT precessions of $I$ and $\varpi$ were simultaneously taken into account by using $k_2^\mathrm{c}=0.228,\,M_\mathrm{c}=1.02\,\mathrm{M}_\odot,\,R_\mathrm{c}=5400\,\mathrm{km},\,\mathcal{I}_\mathrm{c}^{\ton{0}}=1\times 10^{43}\,\mathrm{kg\, m^2}$.}\label{figura3}
\end{center}
\end{figure}
Figure\,\ref{figura3} shows the allowed region in the 3D parameter space $\grf{i_\mathrm{c},\,\zeta_\mathrm{c},\,P_\mathrm{c}}$ determined simultaneously by the constraint on $\dot x_\mathrm{p}^\mathrm{PK}$ of \rfr{boundo}, and by the further condition
\eqi
\left|\dot\varpi^\mathrm{PK}\ton{i_\mathrm{c},\,\zeta_\mathrm{c},\,P_\mathrm{c}}\right|<\upsigma_{\dot\omega_\mathrm{exp}},\lb{peribo}
\eqf
where the error in the periastron precession is assumed to be given by \rfr{erro}. The other system's parameters are as in Figures\,\ref{figura1}-\ref{figura2}.
It is apparent how  the further constraint of \rfr{peribo} may change significantly the picture offered only by \rfr{boundo} and Figure\,\ref{figura1}. Indeed, it restricts the allowed intervals for $\zeta_\mathrm{c}$, and increases the minimum WD's spin period.
\section{Summary and conclusions}\lb{fine}
The recent analysis of the measured secular rate of change $\dot x_\mathrm{p}^\mathrm{exp}$ of the projected semimajor axis $x_\mathrm{p}$ of the pulsar p hosted in the binary system PSR J1141-6545 has been often presented as a successful test of the general relativistic LT effect caused by the angular momentum ${\bds S}^\mathrm{c}$ of the neutron star's companion c, a WD of comparable mass.

In fact, such an interpretation would be valid if all the relevant physical and orbital system's parameters were known independently of the effect itself under consideration. Moreover, even if this were the case, a quantitative measure of the main systematic uncertainty due to the quadrupole mass moment $Q_2^\mathrm{c}$ of the WD should be given. Actually, neither the sizes $S^\mathrm{c},\,Q_2^\mathrm{c}$ nor the orientation in space of its spin axis ${\bds{\hat{S}}}^\mathrm{c}$ are known: several model-dependent and history-dependent assumptions have to be made on the sizes of such key physical parameters. Conversely, one can try to a priori assume the validity of general relativity, and try to constrain some of the still unknown system's parameters like the WD's spin period $P_\mathrm{c}$ and the two angles $i_\mathrm{c},\,\zeta_\mathrm{c}$ of its spin axis.

We did that by studying the regions in the system's 3D parameter space $\grf{i_\mathrm{c},\,\zeta_\mathrm{c},\,P_\mathrm{c}}$ which are allowed by the  constrain that most of the experimental range for the measured $\dot x_\mathrm{p}^\mathrm{exp}$ is due to the sum $\dot x_\mathrm{p}^\mathrm{PK}$ of the PK quadrupolar and LT rates of change.  We found that $P_\mathrm{c}\leq 600\,\mathrm{s}$, with a most probable value around $P_\mathrm{c}\simeq 200-300\,\mathrm{s}$ for a large part  of the allowed values of  $i_\mathrm{c},\,\zeta_\mathrm{c}$. Depending on the value adopted of $P_\mathrm{c}$, it seems that the azimuthal angle $\zeta_\mathrm{c}$ is more tightly constrained than $i_\mathrm{c}$. It is also shown that it is rather unlikely that the systematic bias due to the WD's $Q_2^\mathrm{c}$ can be smaller then $\simeq 1-10\,\mathrm{per\,cent}$ of the LT effect, being, instead, much more likely that it can be as large as $\simeq 30-50\,\mathrm{per\,cent}$ of it. To this aim, it is of great importance the calculation of $Q_2^\mathrm{c}$, which is not a trivial task because of the large uncertainties affecting it.

We also showed that determining the precession $\dot\varpi$ of the longitude of periastron $\varpi$ as a further, independent PK parameter may be of great help in further constraining the WD's parameters of interest. After having figured out a plausible value for the experimental uncertainty $\upsigma_{\dot\varpi_\mathrm{exp}}$ for it, we added the further constraint that it is larger than the theoretically expected PK periastron precession due to both the quadrupole and the LT effect, as claimed by \citet{LTWDPSR20}. The resulting picture in the allowed region of the 3D parameter space turns out to be substantially changed, being the minimum value of $P_\mathrm{c}$ increased and $\zeta_\mathrm{c}$ further restricted.
\section*{Acknowledgements}
I am grateful to an anonymous Editor and an anonymous referee for their helpful comments and remarks.
\bibliography{LTWDPSR}{}

\end{document}